\documentclass[12pt]{article}
\usepackage{amssymb}
\usepackage{epsfig}
\usepackage{float}
\usepackage{graphicx}
\usepackage{amsmath}
\newcommand{\nua}[1]{\ensuremath{\rlap
           {\kern-2.5pt\ensuremath
           {\overset{\scriptscriptstyle(-)}{\phantom{\nu}}}}
           {\ensuremath{{\nu}_{#1}}}}}

\begin{document}

\begin{center}
{\bf Comments on  the determination of the neutrino mass ordering in reactor neutrino experiments}
\end{center}

\begin{center}
S. M. Bilenky
\end{center}
\begin{center}
{\em  Joint Institute for Nuclear Research, Dubna, R-141980,
Russia\\}
{\em TRIUMF
4004 Wesbrook Mall,
Vancouver BC, V6T 2A3
Canada\\}
\end{center}

\begin{abstract}
We consider the problem of determination of the neutrino mass ordering via precise study of
the vacuum neutrino oscillations in the JUNO and other future medium baseline reactor 
neutrino experiments. We are proposing to resolve neutrino mass ordering by determination of the neutrino 
oscillation parameters from analysis of the data of  the  reactor  experiments and 
comparison them with the oscillation parameters obtained from analysis of the solar and KamLAND experiments.
\end{abstract}

The establishment of the character of the neutrino mass ordering (normal or inverted?) is one of the major 
problem of future high precision neutrino oscillation experiments. This problem will be investigated via 
observation of matter effects in the accelerator T2K \cite{Giganti:2015pta} and NOvA 
\cite{Bian:2013saa},  atmospheric  PINGU \cite{Gross:2014oqa} and  ORCA \cite{VanElewyck:2015una} and 
other neutrino experiments. 

The measurement of the angle $\theta_{13}$ in the 
accelerator T2K  \cite{Abe:2015awa} and in the 
reactor Daya Bay \cite{An:2015rpe}, RENO \cite{RENO:2015ksa}  and Double Chooz \cite{Abe:2015rcp} 
experiments opened the way of the determination  of the neutrino mass ordering by the investigation of 
the vacuum neutrino oscillations in reactor neutrino experiments.

Dependence  of the probability of reactor $\nu_{e}$'s to survive on the neutrino mass ordering (NMO)  was  
noticed  in the paper \cite{Bilenky:2001jq}, in which reactor CHOOZ data were analyzed in the framework of 
three-neutrino mixing, and in the paper \cite{Fogli:2001wi}. 

The neutrino mass ordering can be revealed in 
experiments in which effect of both solar and atmospheric mass-squared differences ($\Delta 
m_{S}^{2}$ and $\Delta m_{A}^{2}$) can be observed. It was shown in \cite{Petcov:2001sy,Choubey:2003qx} 
that this condition can be realized in medium baseline reactor neutrino experiments with source-detector 
distance 20-30 km. Later in numerous papers (see 
\cite{Learned:2006wy,Zhan:2008id,Zhan:2009rs,Ghoshal:2010wt,Li:2013zyd,An:2015jdp}) a  
possibility to 
determine   the neutrino mass ordering  in  reactor experiments  was analyzed in details. It was 
established that the optimum baseline is about 50-60 km and energy resolution must be 
$\frac{3\%}{\sqrt{E(MeV)}}$ or better.

Two ambitious  medium baseline reactor neutrino experiments JUNO \cite{An:2015jdp} in China and RENO-50 
\cite{Kim:2014rfa} in Korea were proposed.  The JUNO experiment is now under construction. In this 
experiment 20 kton liquid scintillator detector will be located at the distances $\sim$ 53 km from two 
nuclear power plants. It is planned that the data taking in this experiment will be started  in 2020.

Let us consider  the three-neutrino mixing
\begin{equation}\label{Mix}
    \nu_{lL}=\sum^{3}_{i=1}U_{li} \nu_{iL} \quad (l=e,\mu,\tau),
\end{equation}
where $\nu_{lL}$ ($ l=e,\mu,\tau$) is the flavor neutrino field, $\nu_{i}$ is the field of neutrino (Dirac 
or Majorana)  with mass $m_{i}$  and $U$ is a unitary $3\times 3$ PMNS \cite{Pontecorvo:1957qd,Maki:1962mu} 
mixing matrix which is characterized by three mixing angles  $\theta_{12}, 
\theta_{23},\theta_{13}$ and one $CP$ phase $\delta$. As it is well known, in this case two neutrino mass 
spectra are possible. 

Usually,  neutrino 
masses are labeled in such a way that for both spectra $m_{2}> m_{1}$ and $\Delta m_{12}^{2}=\Delta 
m_{S}^{2}>0$ is solar mass-squared difference.\footnote{We will use the following definition of the 
mass-squared difference: $\Delta m_{ki}^{2}=m_{i}^{2}-m_{k}^{2}.$} 

Possible neutrino mass spectra are determined by the $m_{3}$ mass. There are two possibilities
 \begin{enumerate}
  \item Normal ordering (NO)  $m_{3}>m_{2}>m_{1}$.

\item  Inverted ordering (IO) $m_{2}>m_{1}>m_{3}$.

\end{enumerate}

Following the standard procedure, for the probability of $\bar\nu_{e}$ to survive in vacuum we have
\begin{equation}\label{survival1}
P(\bar\nu_{e}\to \bar\nu_{e})=1 -2\sum_{i>k}|U_{e i}|^{2}|U_{e k}|^{2}(1-\cos2\Delta_{ki}).
\end{equation}
Here
\begin{equation}\label{survival2}
 \Delta_{ki}=\frac{\Delta m^{2}_{ki}L}{4E},
\end{equation}
where $L$ is the distance between neutrino source and detector and $E$ is the neutrino energy. Taking into 
account that 
\begin{equation}\label{survival3}
|U_{e 1}|^{2}=\cos^{2}\theta_{13}\cos^{2}\theta_{12},~~ |U_{e 
2}|^{2}=\cos^{2}\theta_{13}\sin^{2}\theta_{12},~~
|U_{e 3}|^{2}=\sin^{2}\theta_{13}
\end{equation}
we have
\begin{eqnarray}\label{survival4}
&&P(\bar\nu_{e}\to \bar\nu_{e})=1 -\cos^{4}\theta_{13}\sin^{2}2\theta_{12}\sin^{2}\Delta_{S}\nonumber\\
&&-\frac{1}{2}\sin^{2}2\theta_{13}
\left[\cos^{2}\theta_{12}(1-\cos2\Delta_{13})+\sin^{2}\theta_{12}(1-\cos2\Delta_{23})\right].
\end{eqnarray}
Only   last term of this expression, proportional to the small parameter $\sin^{2}2\theta_{13}$, depends on 
the neutrino mass ordering.

In the case of three neutrino masses, there are two independent mass-squared differences. Three neutrino 
mass-squared differences in (\ref{survival4}) (for 
both mass spectra) satisfy the following identity 
\begin{equation}\label{survival6}
\Delta m^{2}_{13}=\Delta m^{2}_{23}+\Delta m^{2}_{12}.
\end{equation}
From (\ref{survival6}) we have
\begin{equation}\label{survival7}
|\Delta m^{2}_{13}|=|\Delta m^{2}_{23}|\pm \Delta m^{2}_{S},
\end{equation}
where $\pm$ corresponds to NO(IO), respectively. Thus, we have
\begin{equation}\label{survival8}
\cos2\Delta_{13}=\cos2|\Delta_{23}|\cos2\Delta_{S}\mp  \sin2|\Delta_{23}|\sin2\Delta_{S}.
\end{equation}

From (\ref{survival4}) and (\ref{survival8}) for the  $\bar\nu_{e}$ survival probability we obtain the 
following expression \cite{Qian:2012xh}
\begin{eqnarray}\label{survival9}
 P(\bar\nu_{e}\to\bar\nu_{e})&=&1 
-\cos^{4}\theta_{13}\sin^{2}2\theta_{12}\sin^{2}\Delta_{S}\nonumber\\
&-&\frac{1}{2}\sin^{2}2\theta_{13} [1-a~\cos(2|\Delta_{23}|\pm \phi)].
\end{eqnarray}
Here
\begin{equation}\label{survival10}
\sin\phi=\frac{1}{a}~\cos^{2}\theta_{12}\sin2\Delta_{S},\quad
\cos\phi=\frac{1}{a}~(\cos^{2}\theta_{12}\cos2\Delta_{S}+\sin^{2}\theta_{12}),
\end{equation}
where
\begin{equation}\label{survival11}
a=\sqrt{1-\sin^{2}2\theta_{12}\sin^{2}\Delta_{12}}
\end{equation}
and $\pm$ correspond to NO and IO, respectively.
 
The expressions (\ref{survival9}), (\ref{survival10}), (\ref{survival11}) were used in  several recent 
papers
(see \cite{Qian:2012xh,Kettell:2013eos,Qian:2015waa,Yang:2015qza}) to  estimate 
the sensitivity of the JUNO experiment to the neutrino mass ordering.\footnote{Similar expression was used 
in \cite{Takaesu:2013wca} for the estimation of the sensitivity of the RENO-50 experiment to NMO.} It was 
suggested in these papers that the only difference between NO and IO
  is the  sign before the phase $\phi$ in the expression (\ref{survival9}).\footnote{See, for example, 
\cite{Kettell:2013eos}: ``As shown in (\ref{survival9}), neutrino mass ordering dependence comes solely 
through the phase shift $\phi$...''} Let us notice, however, that the parameter $|\Delta m^{2}_{23}|$ also 
depends on the neutrino mass ordering.
In fact, we have
\begin{equation}\label{Atm}
|\Delta m^{2}_{23}|= \Delta m_{A}^{2}
~~~\mathrm{NO} ,\quad  |\Delta m^{2}_{23}|=\Delta m_{A}^{2}+\Delta m_{S}^{2}~~~\mathrm{IO},
\end{equation}
where atmospheric mass-squared difference $\Delta m_{A}^{2}$ is determined as follows
\begin{equation}\label{Atm1}
 \Delta m_{A}^{2}=\Delta m^{2}_{23}~~~\mathrm{NO} ,\quad  \Delta m_{A}^{2}= |\Delta m_{13}^{2}| 
~~~\mathrm{IO}.
\end{equation}
From our point of view neutrino oscillation parameters must be determined in a NMO independent way.  
Neutrino mixing angles, $CP$ phase 
and $\Delta m^{2}_{S}$ are determined in such a way. The atmospheric mass-squared difference 
$\Delta m_{A}^{2}$, determined by (\ref{Atm1}), (but not the parameter $|\Delta m^{2}_{23}|$) also satisfies 
this criteria.

From (\ref{survival4}) and (\ref{Atm1}) for the $\bar\nu_{e}$ survival probability  for the 
normal and inverted mass ordering  we have, correspondingly,
\begin{eqnarray}\label{survival14}
&&P^{NO}(\bar\nu_{e}\to \bar\nu_{e})=1 
-\cos^{4}\theta_{13}\sin^{2}2\theta_{12}\sin^{2}\Delta_{S}\nonumber\\
&-&\sin^{2}2\theta_{13}\left[\cos^{2}\theta_{12}\sin^{2}(\Delta_{A}+\Delta_{S})+
\sin^{2}\theta_{12}\sin^{2}\Delta_{A}\right].
\end{eqnarray}
and
\begin{eqnarray}\label{survival14a}
&&P^{IO}(\bar\nu_{e}\to \bar\nu_{e})=1 
-\cos^{4}\theta_{13}\sin^{2}2\theta_{12}\sin^{2}\Delta_{S}\nonumber\\
&&-\sin^{2}2\theta_{13}\left[\cos^{2}\theta_{12}\sin^{2}\Delta_{A}+
\sin^{2}\theta_{12}\sin^{2}(\Delta_{A}+\Delta_{S})\right].
\end{eqnarray}
There are several possibilities to choose NMO independent atmospheric mass-squared difference (see 
\cite{Bilenky:2015bvt}). If $\Delta m_{A}^{2}$ is determined as in \cite{Gonzalez-Garcia:2014bfa}
\begin{equation}\label{NuFit}
\Delta m_{A}^{2}=\Delta m_{13}^{2}~~~\mathrm{NO},\quad  \Delta m_{A}^{2}= |\Delta m_{23}^{2}|~~~\mathrm{IO}
\end{equation}
for the $\bar\nu_{e}$ survival probability in the case of NO and IO we have
\begin{eqnarray}\label{survival15}
&& P^{NO}(\bar\nu_{e}\to\bar\nu_{e})=1 
-\cos\theta_{13}^{4}\sin^{2}2\theta_{12}\sin^{2}\Delta_{S}\nonumber\\
&&-\sin^{2}2\theta_{13}\left[\cos^{2}\theta_{12}\sin^{2}\Delta_{A}+
\sin^{2}\theta_{12}\sin^{2}(\Delta_{A}-\Delta_{S})\right]
\end{eqnarray}
and 
\begin{eqnarray}\label{survival15a}
&& P^{IO}(\bar\nu_{e}\to\bar\nu_{e})=1 
-\cos\theta_{13}^{4}\sin^{2}2\theta_{12}\sin^{2}\Delta_{S}\nonumber\\
&&-\sin^{2}2\theta_{13}\left[\cos^{2}\theta_{12}\sin^{2}(\Delta_{A}-\Delta_{S})
\sin^{2}\theta_{12}\sin^{2}\Delta_{A}\right].
\end{eqnarray}
In \cite{Capozzi:2013psa} the atmospheric mass-squared difference is determined as follows
\begin{equation}\label{Lisi}
\Delta m_{A}^{2}=\frac{1}{2}(\Delta m_{13}^{2}+\Delta m_{23}^{2})~~\mathrm{NO},~~ \Delta m_{A}^{2}= 
\frac{1}{2}|(\Delta m_{13}^{2}+\Delta m_{23}^{2})|~~\mathrm{IO}
\end{equation}
Taking into account that
\begin{equation}\label{Lisi1}
 \Delta m_{13}^{2}=\frac{1}{2}(\Delta m_{13}^{2}+\Delta m_{23}^{2})+\frac{1}{2}\Delta m_{12}^{2},~~
 \Delta m_{23}^{2}=\frac{1}{2}(\Delta m_{13}^{2}+\Delta m_{23}^{2})-\frac{1}{2}\Delta m_{12}^{2}
\end{equation}
for the $\bar\nu_{e}$ survival probability for the normal and inverted ordering we have
\begin{eqnarray}\label{survival16}
&& P^{NO(IO)}(\bar\nu_{e}\to\bar\nu_{e})=1 
-\cos\theta_{13}^{4}\sin^{2}2\theta_{12}\sin^{2}\Delta_{S}\nonumber\\&&
-\sin^{2}2\theta_{13}\left[1-\cos2\Delta_{A}\cos\Delta_{S}\pm 
\sin2\Delta_{A}\sin\Delta_{S}\cos2\theta_{12}\right].
\end{eqnarray}
Let us notice that the atmospheric neutrino mass-squared difference, determined in (\ref{Atm1}) and 
(\ref{NuFit}), 
differ by $\Delta m_{S}^{2}$ and the atmospheric neutrino mass-squared difference, determined in 
(\ref{Lisi}), 
differs  from the atmospheric mass-squared difference determined in (\ref{Atm1}) and (\ref{NuFit}) by 
$\frac{1}{2}\Delta m_{S}^{2}$. In the era of  high presicion neutrino oscillation experiments this 
difference is important. From our point of view a consensus must be found and {\em one universal atmospheric 
neutrino mass-squared difference} have to be determined (see \cite{Bilenky:2015bvt}).

It follows from comparison of (\ref{survival14}) and (\ref{survival14a}) ( (\ref{survival15}) and 
(\ref{survival15a}) and (\ref{survival16})) that 
$P^{NO}(\bar\nu_{e}\to\bar\nu_{e})\leftrightarrows P^{IO}(\bar\nu_{e}\to\bar\nu_{e})$ if we 
change $\cos^{2}\theta_{12}\leftrightarrows \sin^{2}\theta_{12}$. This is a natural symmetry of the 
expressions for $\bar\nu_{e}$ transition probabilitities if 
the atmospheric mass-squared difference is determined in NMO independent way (see  
\cite{Bilenky:2001jq}). This symmetry suggest a possible  way for the 
establishment of the neutrino mass ordering. Let us assume that atmospheric mass-squared difference was 
determined as in (\ref{Atm}). We can rewrite the equations
(\ref{survival14}) and (\ref{survival14a}) in the form
\begin{eqnarray}\label{survival17}
&&P(\bar\nu_{e}\to \bar\nu_{e})=1 -\cos^{4}\theta_{13}\sin^{2}2\theta_{12}\sin^{2}\Delta_{S}\nonumber\\
&&-\sin^{2}2\theta_{13}\left[(1-X)\sin^{2}(\Delta_{A}+\Delta_{S})+
X\sin^{2}\Delta_{A}\right].
\end{eqnarray}
Here
\begin{equation}\label{survival18}
X=\sin^{2}\theta_{12}~(\mathrm{NO});~~ X=\cos^{2}\theta_{12}~(\mathrm{IO}).
\end{equation}
The value of the parameter $\tan^{2}\theta_{12}$ is known from analysis of the data of the KamLAND and solar 
neutrino experiments. From the latest three-neutrino analysis of the data it was found 
\cite{Shimizu:2015nxp}
\begin{equation}\label{survival19}
\tan^{2}\theta_{12}=0.437^{+0.029}_{-0.026}
\end{equation}
Thus, in order to reveal the neutrino mass ordering  we need to determine from analysis of the data of 
future medium baseline reactor neutrino experiments the parameter $X$  and to check 
whether $X$ is equal to
\begin{equation}\label{survival20}
\sin^{2}\theta_{12}=0.304\pm 0.014 
\end{equation}
or
\begin{equation}\label{survival21}
\cos^{2}\theta_{12}=0.696\pm 0.014.
\end{equation}
In the first case neutrino mass ordering is the normal one and in the second it is the inverted one.
 
It was shown in \cite{Kettell:2013eos,An:2015jdp} that after six years of data 
taking the parameters $\sin 2\theta_{12}$, $\Delta m_{S}^{2}$ and $\Delta m_{A}^{2}$
 will be determined in the JUNO experiment with accuracy better than 1\%. From the 
latest result of the Daya Bay experiment it was found \cite{Tang:2015vug}
\begin{equation}\label{survival22}
\sin^{2}2\theta_{13}=0.084 \pm 0.005.
\end{equation}
Apparently, the future high precision  JUNO and RENO-50 experiments and future knowledge of the 
neutrino oscillation parameters 
will allow to distinguish the values in (\ref{survival20}) and (\ref{survival21}).

I am thankful to C. Giunti for numerous fruitful discussions. I acknowledge the support of RFFI grant 
16-02-01104.

\end{document}